\documentclass[a4paper,10pt]{article}
\usepackage[utf8x]{inputenc}

\title{ Boundary Effects in the BLG Theory  }
\author{ Mir Faizal \\
Mathematical Institute, University of Oxford
\\ Oxford
OX1 3LB, United Kingdom 
 }
\date{}

\begin{document}

\maketitle

\begin{abstract} In this paper we will analyse a system of multiple M2-branes 
in between two M5-branes. This will be analysed by studying Bagger-Lambert-Gustavsson (BLG) theory 
on a manifold with two boundaries. The original BLG theory will be modified to make it 
 gauge and supersymmetric invariant  in presence of the boundaries. However, this modified theory will 
 only preserve half the supersymmetry 
of the original theory. 
We will also analyse the deformation of this theory caused by noncommutativity
between Grassman coordinates and spacetime coordinates. 
Finally, we will  analyse the Higgsing of this theory to deformed D2-branes with boundaries. 
\end{abstract}

\section{Introduction}
 BLG theory is the theory of multiple M2-branes and it is 
 constructed using  the  Lie $3$-algebra \cite{1}-\cite{5}.  
The BLG theory has been analysed 
in the   $\mathcal{N} = 1$ superspace  
formalism \cite{14}-\cite{ab1}.  
It may be noted that the dimensionally reduction of the multiple M2-branes in 
$\mathcal{N}=1$ superspace formalism has also been analysed \cite{dr}.
 In this theory a map to a Green-Schwarz 
string wrapping a nontrivial circle in $C^4/Z_k$ has also been 
constructed. The mass deformation the BLG theory has also been studied \cite{mass1}-\cite{mass2}.
 This mass-deformed theory  preserves maximal supersymmetry but is not conformal. In fact, this mass 
deformed BLG theory also  has a maximally supersymmetric fuzzy two-sphere vacuum
solution in which the scalar fields are proportional to the $TGRVV$ matrices \cite{tg}. 
Fluctuations about  fuzzy two-sphere in this theory can be described by D4-branes. It is expected that 
this corresponded to the dimensional reduction of a M5-brane  \cite{m5}-\cite{m51}.

There is a 
duality between M-theory and $II$ 
string theory. So,  a 
 deformation of the supersymmetric  algebra on the string 
theory side will 
correspond to some deformation of this algebra on the
  M-theory side also.  
In analogy with the deformation of D-branes by two-form  fields 
a deformation of M-theory can  occur due to three-form fields 
which occur 
naturally in M5-branes. The coupling of BLG theory to three form fields 
can be useful in describing the physics of M2-branes ending on 
M5-branes as M5-branes in M-theory act 
as analogous objects 
to a D-brane in string theory, in the sense that M2-branes  can end
 on them. 
The action for a single 
M5-brane  has in fact been derived  by demanding
the $\kappa$-symmetry of the open membrane ending on it \cite{14om}.
Even though the action for a single M5-brane
 is known, the  
action   for multiple M5-branes is not
known \cite{41s}-\cite{42s}. 
 Thus, 
the analysis of 
BLG theory on boundary coupled to a background three-form field strength might 
give some useful insight in deriving the action for multiple M5-branes.

Apart from these backgrounds string theory has been studied in graviphoton       background. 
The deformation caused by this graviphoton background has also been analysed  \cite{gravity}-\cite{gravity1}.
In fact, compactification of  open string amplitudes with the D3-branes in type $IIB$ superstring theory    on $C2/Z2$ has also
been studied \cite{gp}-\cite{gp1}. This has been done by introducing a constant graviphoton background along
the branes and calculate disk amplitudes using the $NSR$ formalism. In doing so a 
zero slope limit was   taken  the effective Lagrangian on the D3-branes deformed
by the graviphoton background was investigated. Furthermore, $\mathcal{N}=1$    supersymmetric 
gauge theory, with chiral matter multiplets in the fundamental representation of the gauge group,  deformed by 
 graviphoton background has also been analysed \cite{gp2}.
In doing so the perturbation theory scheme of computing these correlation functions has also been studied. 
The relations between all the vacua of Lorentzian and Euclidean  SUGRAs in various dimensions  with $8$ supercharges, 
finding a new limiting procedure that takes us from the over-rotating near-horizon BMPV black hole to the Godel spacetime have also 
been analysed 
\cite{gp4}.  
The timelike compactification of the maximally supersymmetric Godel solution of $\mathcal{N}=1$ SUGRA in five dimensions gives a maximally supersymmetric
 solution of pure Euclidean $\mathcal{N}=2$ theory in four dimensions  with flat space but non-trivial anti-self dual vector field flux 
 that
 can be interpreted as an $U(1)$ instanton on the $4$-torus and   it coincides with the graviphoton background.
 In this background no supersymmetry is broken. 
As we want to retain high amount of supersymmetry for the BLG theory, so we will analyse it in this background. 
In this paper we will thus analyse the deformed BLG theory 
in $\mathcal{N} =1$ superspace formalism. The deformation will be caused by a graviphoton       background.

As M2-branes can end on M5-branes, so we need to analyse the BLG theory in presence of a boundary 
In a supersymmetric theory, the presence of a boundary  breaks  the supersymmetry.  The boundary
obviously breaks translational symmetry and since supersymmetry closes on translations,
 it is inevitable that the presence of boundary will also break supersymmetry.
However, half of the the supersymmetry can
be preserved by adding a boundary term to the bulk action, such that 
the supersymmetric variation of this boundary term 
exactly cancels the boundary piece generated by the 
supersymmetric transformation of the bulk action \cite{boundary}-\cite{boundary1}. 
This has been used for analysing the ABJM theory \cite{faizal}-\cite{faizal2}  and the BLG theory \cite{faizal1}
with one boundary. 
Here we shall first generalize it to case where two boundaries are present. 
In the case of ABJM theory and BLG theory  with a single boundary, we had to add one new bulk field. 
In this paper we will show that even for the case of two boundaries we only need to add one 
new bulk field. We will also analyse the deformation of the theory thus obtained by a graviphoton 
background and finally analyse the Higgsing of the M2-brane with boundary to D2-brane with boundary.

\section{ BLG Theory }
In this section we will review the construction of a gauge and supersymmetric invariant BLG theory 
on a manifold with  boundaries. 
The BLG theory is based on gauge symmetries generated by gauge fields which take values in a Lie $3$-algebra,
$
[T^A,T^B,T^C] = f^{ABC}_D T^D,
$
were $f^{ABC}_D$ are the structure constants and 
 $T^A$ are the generators of this Lie $3$-algebra with $h_{AB} = Tr (T_A T_B) $.  
 These structure constants are totally antisymmetric 
  in $A,B,C$ and  satisfy the Jacobi identity, 
$
f^{[ABC}_G f^{D]EG}_H = 0
$.
It is useful to define  \cite{blgblg}
$
C^{AB,CD}_{EF} = 2f^{AB[C}_{[E} \delta^{D]}_{F]}, 
$
which  are anti-symmetric in the  pair of indices $AB$ and $CD$ and also satisfy the Jacobi identity,
$
C^{AB,CD}_{EF} C^{GH,EF}_{KL} + C^{GH,AB}_{EF} C^{CD,EF}_{KL} +C^{CD,GH}_{EF} C^{AB,GH}_{KL} =0.
$
The BLG theory  has $\mathcal{N}  =8$ supersymmetry. However, we will write it in $\mathcal{N} = 1$ superspace formalism with 
 manifest $\mathcal{N} =1$ supersymmetry generated by the supercharge,  
$
 Q_a = \partial_a - (\gamma^\mu \partial_\mu)^b_a \theta_b
$,
which commutes with the super-derivative 
$
  D_a = \partial_a + (\gamma^\mu \partial_\mu)^b_a \theta_b
$.
Now  the Lagrangian for the BLG theory can be written as 
\begin{equation}
  \mathcal{L}  = -\nabla^2 [\mathcal{CS}(\Gamma) + \mathcal{M} (X^I, X^{\dagger I }) ]_|
\end{equation}
where
\begin{eqnarray}
\mathcal{CS} (\Gamma) &=& \frac{k}{4\pi}f^{ABCD}\Gamma^{a}_{ AB}  \Omega_{a CD}, 
\nonumber \\ 
\mathcal{M} ( X^I, X^{\dagger I }) &=& 
\frac{1}{4}  (\nabla^a X^I)^A  (\nabla_a X^{\dagger}_I)_A \nonumber \\  && -\frac{2\pi}{k}
\epsilon_{IJKL} f^{ABCD}X^I_A X^{K \dagger}_B X^{J}_C  X^{L \dagger}_D, 
\end{eqnarray}
Here  $X^I_A, X^{\dagger I}_A $ are scalar superfields,  $\Gamma_{AB}^{a}$ is a spinor gauge field and 
\begin{eqnarray}
 \Omega_{ a AB} & = & \omega_{a AB} - \frac{1}{3}C^{CD,EF}_{AB}[\Gamma^{b CD}, \Gamma_{ab EF}] \\
 \omega_{a AB} & = & \frac{1}{2} D^b D_a \Gamma_{b AB} -i  C^{CD,EF}_{AB}[\Gamma^b_{CD} , D_b \Gamma_{a EF}] \nonumber \\ && -
 \frac{1}{3}C^{CD,EF}_{AB} C^{GH,IJ}_{EF}[ \Gamma^b_{CD} ,
\{ \Gamma_{b GH} , \Gamma_{a IJ}\} ],  \\
 \Gamma_{ab AB} & = & -\frac{i}{2} \left[ D_{(a}\Gamma_{b) AB} 
- 2 i C^{CD,EF}_{AB}\{\Gamma_{a CD}, \Gamma_{b EF}\} \right]. 
\end{eqnarray}  
The covariant derivatives of these fields are defined as 
\begin{eqnarray}
  \nabla_a  X^I_A&=& D_a X_A -i f^{BCD}_A \Gamma_{a BC} X^{I}_{ D},\nonumber\\
 \nabla_a X^{I \dagger}_A &=& D_a X^{I \dagger}_A + if^{BCD}_A X^{I \dagger}_D \Gamma^{a BC}, \\
( \nabla_a \Gamma_b)_{AB} &=& D_a \Gamma_{b AB} + C^{CD,EF}_{AB}\Gamma_{CD a} \Gamma_{b EF}. 
\end{eqnarray}
The covariant derivative of $\omega_{a AB}$ vanishes, $\nabla^a \omega_{a AB} =0$.
Now let us consider the  gauge transformations generated by $ u = \exp ( i \Lambda^{AB} T_A T_B),$
\begin{eqnarray}
    \Gamma_{a } \rightarrow  i u \, \nabla_a u^{-1}, &&
X^{I }\rightarrow  u X^I,\nonumber\\
 X^{I  \dagger} \rightarrow X^{I \dagger}_Bu^{-1},&&
\end{eqnarray}
where 
$X^I = X^I_A T^A, \, X^{I \dagger} =  X^{I \dagger}_AT^A, \, 
 \Gamma_{a } = \Gamma_{a AB} T^A T^B $. Under these gauge transformations the BLG Lagrangian transforms as 
\begin{eqnarray}
  \delta \mathcal{L}  &=& 
  \mathcal{L} (\mathcal{CS}(\Gamma^u) + \mathcal{M} (X^{I u}, X^{\dagger I u}))  
-  \mathcal{L} (\mathcal{CS}(\Gamma) + \mathcal{M} (X^I, X^{\dagger I })) \nonumber \\ 
&=& \mathcal{D}_{\mu} [\Psi (\gamma^\mu, \Gamma, X^I, X^{I \dagger})]_|,
\end{eqnarray}
where $\Gamma_a^u, X^{I u}, X^{\dagger I u}$ denoted the gauge transformation of these fields by $u$. 
So, when no boundaries are present the BLG theory is gauge invariant $\delta  \mathcal{L}  =0$. 
 It is also transforms under  the supersymmetry transformations generated by $Q_a$ as
\begin{eqnarray}
 \delta_{S}  \mathcal{L}_{BLG} &=&   \epsilon^a Q_a  \mathcal{L}_{BLG} \nonumber \\ 
&=&  \mathcal{D}_{\mu} [\Phi (\gamma^\mu, \Gamma, X^I, X^{I \dagger})]_|. 
\end{eqnarray}
So, when no boundaries are present the BLG theory is also invariant under this $\mathcal{N} =1$ supersymmetry, 
 $\delta_S  \mathcal{L}  =0$. 
\section{Boundary Effects}
The BLG Lagrangian is invariant under supersymmetric and gauge transformations as under both these transformations, 
the Lagrangian density transforms into a total derivative which vanishes in absence of a boundary. 
However,  if we have M2-branes between two M5-branes, then from the M2-brane perspective, the system will be described by 
M2-branes with two boundaries. BLG theory with one boundary condition has been already studied  \cite{faizal1}. 
Thus, if the two M5-branes are placed at $x_3=c_1$ and $x_3=c_2$, where $c_1$ and $c_2$ are constants, 
 then M2-branes between them will be described by BLG theory with two boundaries. 
 Both the gauge and supersymmetric transformations of the BLG Lagrangian will generate 
boundary terms corresponding to them. 
However, it is possible to modify the original BLG theory to obtain a gauge invariant theory which preserves half 
the supersymmetry in presence of these boundaries. 
We  denoted the induced value of the fields $X, X^{\dagger},\Gamma_a, \Lambda$ on the boundary $x_3 =c_1$ as 
 by $X_1, {X^{\dagger}}_1,\Gamma_{a1}, \Lambda_{1}$ and 
the induced value of the fields $X, X^{\dagger},\Gamma_a, \Lambda$ on the boundary $x_3 =c_2$ as 
 by $X_2, {X^{\dagger}}_2,\Gamma_{a2}, \Lambda_{2}$. We also denote the 
 induced value of the super-derivative $D_a$ and the 
super-covariant derivative $\nabla_a$ on the boundaries as 
$D_{a}'$ and $\nabla_{a}'$, respectively.
Now we define projection operators $P_{\pm}$ as
$ (P_{\pm  })_{ab}  =  (C_{ab} \pm (\gamma^{3})_{ab})/2$,
and so the generator of $\mathcal{N} =1$ supersymmetry can  expressed  as
$
  \epsilon^a Q_a = \epsilon^{a -} Q_{- a} + \epsilon^{ a + } Q_{+ a} $.
In presence of these boundaries we can only preserve the supersymmetry generated by  $Q_{a-}$ or $Q_{a+}$, but not both of them. 
Furthermore, to make the Lagrangian gauge invariant we add extra degrees of freedom $v$, which transform as
\begin{equation}
 v \to v u^{-1},
\end{equation}
where $ u = \exp ( i \Lambda^{AB} T_A T_B)$.
We  let $v_1$  and $v_2$ be the the induced values of $v$ on the boundaries $x_3 = c_1$ and $x_3 = c_2$, respectively.
 We also define $\mathcal{D}_\mu $ to be  the ordinary  covariant derivative. 
Now the gauge invariant Lagrangian that is invariant 
under the supersymmetric transformations generated by  $Q_{a+}$ can be written as
\begin{eqnarray}
 \mathcal{L}_{sg+}&=& - \nabla_{+}' [\mathcal{CS}^+(\Gamma^v) + \mathcal{M}^+ (X^{I }, X^{\dagger I }) \nonumber \\ 
&& +\mathcal{K}_+ (\Gamma_1, v_1)+  \mathcal{K}_+ (\Gamma_2, v_2 ) ]_{\theta_+ =0},
\end{eqnarray}
where $\Gamma^v_a$ denote the gauge transformation of $\Gamma_a$ by $v$ and 
\begin{eqnarray}
\mathcal{CS}^+(\Gamma^v) &=& \nabla_-  [\mathcal{CS}(\Gamma^v) ]_{\theta_- =0}, \nonumber \\ 
\mathcal{M}^+ (X^{I }, X^{\dagger I })&=& 
\nabla_- [ \mathcal{M} (X^{I }, X^{\dagger I }) ]_{\theta_- =0},\nonumber \\ 
\mathcal{K}_+ (\Gamma_1, v_1 )&=&  -\frac{k}{2\pi} 
[ f_{ABCD}
({v}_1^{-1} \nabla_{+}' v_1)^{AB} ({v}_1^{-1} \mathcal{D}_{-}' v_1)^{CD}], \nonumber \\ 
\mathcal{K}_+ (\Gamma_2, v_2 )&=&- \frac{k}{2\pi} 
[ f_{ABCD}
({v}_2^{-1} \nabla_{+}' v_2)^{AB} ({v}_2^{-1} \mathcal{D}_{-}' v_2)^{CD}].
\end{eqnarray}
It may be noted that
 $\mathcal{S}^+(\Gamma_1, v_1) + \mathcal{S}^+(\Gamma_2, v_2) =\mathcal{CS}(\Gamma^v) - \mathcal{CS}(\Gamma)$ is the boundary potential.  So, 
$\mathcal{CS}(\Gamma^v) =\mathcal{CS}(\Gamma) +  \mathcal{S}^+(\Gamma_1, v_1) + \mathcal{S}^+(\Gamma_2, v_2) $ is the total
potential of the theory. In case there is no coupling to the bulk fields this reduces to a potential term
 for two Wess-Zumino-Witten models,
\begin{eqnarray}
\nabla_{+}' \mathcal{S}^+(\Gamma_1, v_1) &=& -\frac{k}{2\pi} {\nabla}_{+} 'C^{CD, EF}_{AB} \left[ [(v_1^{-1} \mathcal{D}_{-}' v_{1})^{AB},  
(v^{-1}_1 \mathcal{D}'_3 v)_{CD}]\right. \nonumber \\ && \left. \times
(v^{-1}_1 \nabla_{+}' v_1)_{EF} \right]_|, \nonumber \\
\nabla_{+}' \mathcal{S}^+(\Gamma_2, v_2) &=& -\frac{k}{2\pi} {\nabla}_{+}'C^{CD, EF}_{AB} \left[ [(v^{-1}_2 \mathcal{D}_{-}' v^{2})^{AB},  
(v^{-1}_2 \mathcal{D}'_3 v_2)_{CD}]\right. \nonumber \\ && \left. \times
(v^{-1}_2 \nabla_{+}' v_2)_{EF} \right]_|. 
\end{eqnarray}
 So,  
here it can be viewed as the potential term  some gauged Wess-Zumino-Witten models.
Similarly, we can show that the  gauge invariant Lagrangian that is invariant 
under the supersymmetric transformations generated by  $Q_{a-}$ can be written as
\begin{eqnarray}
 \mathcal{L}_{sg-}&=& - \nabla_- '[\mathcal{CS}^-(\Gamma^v) + \mathcal{M}^- (X^{I }, X^{\dagger I }) \nonumber \\ 
&&+  \mathcal{K}_- (\Gamma_1, v_1) + \mathcal{K}_+ (\Gamma_2, v_2 ) ]_{\theta_- =0},
\end{eqnarray}
where 
\begin{eqnarray}
\mathcal{CS}^-(\Gamma^v) &=& \nabla_+  [\mathcal{CS}(\Gamma^v) ]_{\theta_+ =0}, \nonumber \\ 
\mathcal{M}^- (X^{I }, X^{\dagger I })&=& 
\nabla_+ [ \mathcal{M} (X^{I }, X^{\dagger I }) ]_{\theta_+ =0},\nonumber \\ 
\mathcal{K}_- (\Gamma_1, v_1 )&=& - \frac{k}{2\pi} 
[ f_{ABCD}
({v}_1^{-1} \nabla_{-}' v_1)^{AB} ({v}_1^{-1} \mathcal{D}_{+}' v_1)^{CD}], \nonumber \\ 
\mathcal{K}_- (\Gamma_2, v_2 )&=& -\frac{k}{2\pi} 
[ f_{ABCD}
({v}_2^{-1} \nabla_{-}' v_2)^{AB} ({v}_2^{-1} \mathcal{D}_{+}' v_2)^{CD}].
\end{eqnarray}
Here again $ \mathcal{CS}(\Gamma^v) = \mathcal{CS} (\Gamma) + \mathcal{S}^- (\Gamma_1, v_1) + \mathcal{S}^- (\Gamma_2, v_2)$ is the  total
potential of the theory, which is given by the sum of two gauged Wess-Zumino-Witten models with the Chern-Simons term. 
When there is coupling to the bulk fields the gauged Wess-Zumino-Witten model reduced to 
\begin{eqnarray}
\nabla_{-}' \mathcal{S}^-(\Gamma_1, v_1) &=& -\frac{k}{2\pi}  C^{CD, EF}_{AB}{\nabla}_{-}' \left[ [(v_1^{-1} \mathcal{D}_{+}' v_{1})^{AB},  
(v^{-1}_1 \mathcal{D}_3' v)_{CD}]\right. \nonumber \\ && \left. \times
(v^{-1}_1 \nabla_{-}' v_1)_{EF} \right]_|, \nonumber \\
\nabla_{-}' \mathcal{S}^-(\Gamma_2, v_2) &=& -\frac{k}{2\pi}C^{CD, EF}_{AB} {\nabla}_{-}'  \left[ [(v^{-1}_2 \mathcal{D}_{+}' v^{2})^{AB},  
(v^{-1}_2 \mathcal{D}'_3 v_2)_{CD}]\right. \nonumber \\ && \left. \times
(v^{-1}_2 \nabla_{-}' v_2)_{EF} \right]_|. 
\end{eqnarray}
We have derived a gauge and supersymmetric invariant theory of M2-branes placed in between two M5-branes. 
It may be noted that we only needed to introduce one bulk field $v$ to construct a gauge invariant theory, even in presence of two boundaries. 

\section{  Deformation }
In M-theory there is a three form field $C^{\sigma\nu\tau}$  . Now if $H^{\sigma\nu\tau\rho}  $ is the field strength of 
the $C$, then  deformations of the super-algebra can be caused by 
graviphoton can be caused by $(H_{\sigma\nu\tau\rho}\gamma^\sigma\gamma^\nu\gamma^\tau\gamma^\rho)^{ab} \phi_{a+}^\mu$ and 
$(H_{\sigma\nu\tau\rho}\gamma^\sigma\gamma^\nu\gamma^\tau\gamma^\rho)^{ab} \phi_{a-}^\mu$. 
The graviphoton background  leads to the following deformation, 
$ 
 {[\theta^+, y^\mu ]} = C^{+\mu} $ and $
 {[\theta^-, y^\mu ]} = C^{-\mu}
$.
These deformation induces the following star products between fields  \cite{gravity}-\cite{gravity1}
\begin{eqnarray}
 X^{I \dagger}(\theta, y) \star^+   X_{I }(\theta, y) &=& E^+
 X^{I \dagger}(y_1,\theta_1) X_{I }  (y_2, \theta_2)
\left. \right|_{y_1=y_2=y, \; \theta_1=\theta_2=\theta},\nonumber \\ 
 X^{I \dagger}(\theta, y) \star^-   X_{I }(\theta, y) &=& E^-
 X^{I \dagger}(y_1,\theta_1) X_{I }  (y_2, \theta_2)
\left. \right|_{y_1=y_2=y, \; \theta_1=\theta_2=\theta},
\end{eqnarray}
where 
\begin{eqnarray}
 E^+ &=& \exp -\frac{i}{2} \left(
 C^{+ \mu } (\partial^{+2}  \partial^1_\mu - \partial^2_\mu \partial^{+1} ) \right), \nonumber \\ 
 E^- &=& \exp -\frac{i}{2} \left(
 C^{-\mu } (\partial^{-2} \partial^1_\mu - \partial^2_\mu \partial^{-1} ) \right).
\end{eqnarray}
The only known example of the Lie $3$-algebra is  $h_{AB} = \delta_{AB}$ and $f_{ABCD} =  \epsilon_{ABCD}$
\cite{three}-\cite{three1}. The $SO(4)$ symmetric of this theory can be decomposed into 
$SU(2) \times SU(2)$. The bulk fields now get subtable contacted with the 
generators of the $SU(2)$ Lie algebra. If $t_\alpha$ are the generators of the $SU(2) $ lie algebra, 
 $[t_\alpha, t_\beta] = i f_{\alpha \beta }^\gamma t_\gamma$, then 
$X^I = X^{I\alpha} t_\alpha, X^{\dagger I \alpha} t_{\alpha},\Gamma^{a} =  \Gamma^{\alpha a} t_{\alpha}, 
\Omega = \Omega ^\alpha t_\alpha, v = v ^\alpha t_\alpha,
 \tilde \Gamma^{a} =  \Gamma^{\alpha a} t_{\alpha}, 
 \tilde \Omega =  \tilde \Omega ^\alpha t_\alpha,  \tilde v =  \tilde v ^\alpha t_\alpha, $. 
Thus, the Lagrangian for the Chern-Simons on manifold without boundaries will be given by 
 $\mathcal{L}(\Gamma)_\pm = \nabla^2 [\Gamma^a \star^\pm\Omega_a]_{|} $
 and $\mathcal{L}  (\tilde \Gamma)_\pm =\nabla^2 [\tilde\Gamma^a \star^\pm \tilde\Omega_a]_{|} $.
So, under this decomposition the bulk theory can be written as 
\begin{eqnarray}
   \mathcal{L}_{sg+}  &=& -\nabla_+' [\mathcal{C}^{+}(\Gamma^v)_{\star \pm} - \mathcal{C}^+(\tilde \Gamma^v)_{\star \pm}  
+ \mathcal{MA}^+ (X^I, X^{\dagger I }) _{\star \pm} 
\nonumber \\ && 
+{K}_+ (\Gamma_1, v_1)_{\star \pm}  +  {K}_+ (\Gamma_2, v_2 )_{\star \pm}   \nonumber \\ &&
+  {K}_+ (\tilde \Gamma_1, \tilde v_1)_{\star \pm}  + {K}_+ 
(\tilde \Gamma_2, \tilde v_2 )_{\star \pm}  ]_{\theta_+ =0},
\nonumber \\ 
   \mathcal{L}_{sg-}  &=& -\nabla_-' [\mathcal{C}^{-}(\Gamma^v)_{\star \pm}  - \mathcal{C}^-(\tilde \Gamma^v)_{\star \pm} 
 + \mathcal{MA}^- (X^I, X^{\dagger I })_{\star \pm}  
\nonumber \\ &&
+{K}_- (\Gamma_1, v_1)_{\star \pm}  +{K}_- (\Gamma_2, v_2 )_{\star \pm} 
 \nonumber \\ &&
+{K}_- (\tilde \Gamma_1, \tilde v_1)_{\star \pm}  +K_- (\tilde \Gamma_2, \tilde v_2 )_{\star \pm}  ]_{\theta_- =0}. 
\end{eqnarray}
Here $\mathcal{MA} (X^I, X^{\dagger I })_{\star \pm} $ is Lagrangian for the 
 matter fields and    covariant derivatives for the matter fields  are now given by 
$ \nabla_a  X^I= D_a X -i  \Gamma_{a }\star^\pm X^{I} + i  X^{I} \star^\pm \tilde \Gamma_{a}$ and 
$\nabla_a X^{I \dagger} = D_a X^{I \dagger} + i \tilde \Gamma^{a}\star^\pm X^{I \dagger}  - X^{I}\star^\pm  \tilde \Gamma_{a}$.  
The kinetic terms for the boundary theory are $K_{\pm} (\tilde \Gamma_1, \tilde v_1)_{\star \pm} , K_{\pm}(\tilde \Gamma_2, \tilde v_2)_{\star \pm}  $ and $ 
K_{\pm} ( \Gamma_1,  v_1)_{\star \pm} , K_{\pm}( \Gamma_2,  v_2)_{\star \pm} $ . 
It may be noted that the boundary potential now becomes 
 $\mathcal{SB}^+(\Gamma_1, v_1)_{\star \pm}  + \mathcal{SB}^+(\Gamma_2, v_2)_{\star \pm} 
-\mathcal{SB}^+(\tilde \Gamma_1, \tilde v_1)_{\star \pm} -\mathcal{SB}^+(\tilde \Gamma_2, \tilde v_2)_{\star \pm} 
 =\mathcal{C}(\Gamma^v)_{\star \pm}  - \mathcal{C}(\Gamma)_{\star \pm}  -\mathcal{C}(\tilde \Gamma^v)_{\star \pm} 
 + \mathcal{C}(\tilde \Gamma)_{\star \pm} $.
 So, the total potential for the theory is given by 
$\mathcal{C}(\Gamma^v)_{\star \pm}  - \mathcal{C}(\tilde \Gamma^{\tilde v} )_{\star \pm} $ and 
thus the projections of this total potential term are given by 
$\mathcal{C}^\pm(\Gamma^v)_{\star \pm} $ and $\mathcal{C}^\pm (\tilde\Gamma^{\tilde v })_{\star \pm} $.  

\section{Higgsing}
Now if one of the scalar fields is given a vacuum expectation value, $\langle X \rangle  = \mu \neq 0$,  then the 
the symmetry  group $SU(2) \times SU(2)$ is  spontaneously broken to   its diagonal
subgroup $SU(2)$ \cite{higg}. Now if $A_a = (\Gamma_a - \tilde \Gamma_a)/2$ is the superfield associated with the broken gauge group
and $B_a = (\Gamma_a + \tilde \Gamma_a)/2$ is the superfield  associated with the unbroken gauge group, then the action for 
the BLG theory  on manifolds without a boundary becomes,
\begin{equation}
   \mathcal{L}_{ym}  = - \frac{1}{g^2}\nabla^2  [W^a\star^\pm  W_a + \nabla^a\star^\pm  X^{I\dagger}\star^\pm \nabla_a\star^\pm X^I + 
P[X^{I\dagger}, X^I]_{\star^\pm} ]_|, 
\end{equation}
where $g = 2\pi \nu k^{-1}$, $P[X^{I\dagger}, X^I]_{\star^\pm} $ is the potential term obtained after eliminating $A_a$, 
  and $W_a$ is the field strength given by 
\begin{eqnarray}
  W_a &=& \frac{1}{2} D^b D_a B_b - \frac{i}{2} 
 [B^b , D_b B_a]_{\star^\pm}    -
 \frac{1}{6} [ B^b ,
\{ B_b , B_a\}_{\star^\pm}   ]_{\star^\pm}.
\end{eqnarray}
Here the covariant derivatives are given by $ \nabla_a    X^I =  D_a X^I -i B_a  \star^\pm   X^I
$ and $\nabla_a    X^{I\dagger} = D_a X^{I\dagger} + i B_a  \star^\pm   X^{I\dagger}$. 
It is possible to keep
$g$ fixed in   the limit $\nu \to \infty$ and $k \to \infty$ and so we 
have only considered the leading order terms in powers of $\nu$ and $k$.
 Now if the finite gauge transformations generated by $SU(2)\times SU(2)$ are denoted by $q$ and $\tilde q$ and  $ p = q + \tilde q$, 
  then full finite  gauge 
transformations  under which this theory is invariant  are given by
\begin{eqnarray}
 B_a \rightarrow i p {     }\star^\pm \nabla_a \star^\pm{     } p^{-1},&&
 X^I \rightarrow  p  {     }\star^\pm X^I ,\nonumber \\ 
 X^{I\dagger} \rightarrow   X^{I\dagger} {     }\star^\pm p^{-1}, &&   W_a = p  \star^\pm  W_a \star^\pm    p^{-1}.
\end{eqnarray}
This is thus a super-Yang-Mills theory that occurs as a low energy approximation to D2-brane action. 
Now in presence of  boundaries the super-Yang-Mills theory is still gauge invariant. 
So, in presence of a boundary we have to only take care of the supersymmetry. Thus, we can write 
Lagrangian that preserves half of the supersymmetry as follows, 
$
 \mathcal{L}_{yms + } =-\nabla_+' [\mathcal{Y}^+_{\star \pm} ]_{\theta_+ =0}/ g^2,
$ and $ 
 \mathcal{L}_{yms - } =-\nabla_-'[\mathcal{Y}^-_{\star \pm} ]_{\theta_- =0}/g^2,
$
where 
\begin{eqnarray}
\mathcal{Y}^+ _{\star \pm} &=& \nabla_- [  \nabla^a\star^\pm X^{I\dagger} \star^\pm\nabla_a\star^\pm X^I 
\nonumber \\  && + W^a \star^\pm W_a + P [X^{I\dagger}, X^I]_{\star^\pm} ]_{\theta_+ =0},
\nonumber \\
\mathcal{Y}^-_{\star \pm} &=& \nabla_+ [ \nabla^a \star^\pm X^{I\dagger} \star^\pm \nabla_a\star^\pm X^I 
\nonumber \\  && + W^a \star^\pm W_a + P[X^{I\dagger}, X^I]_{\star^\pm} ]_{\theta_- =0},
\end{eqnarray}
Now if we start from a BLG theory with boundaries and   again set  $\langle X \rangle  = \mu \neq 0$, spontaneously breaking the 
the symmetry  group $SU(2) \times SU(2)$   its diagonal
subgroup $SU(2)$, then we get, $ 
  \mathcal{L}_{m + } =  \mathcal{L}_{yms +  }  + \mathcal{L}_{z +  } $ and $  
 \mathcal{L}_{m - } =\mathcal{L}_{yms -  }  + \mathcal{L}_{z -  },
$ where  
$ 
 \mathcal{L}_{z + } = - \nabla_+[  \mathcal{Z}^+{(B, v)}_{\star \pm} ]_{\theta_+ =0} $ and $ 
 \mathcal{L}_{z - } = - \nabla_-[ \mathcal{Z}^-{(B,   v)}_{\star \pm}]_{\theta_- =0}  
$. Here
 $Z^{\pm} (B, v)_{\star \pm} =0$, when $v =0$. Now as $ \mathcal{L}_{yms \pm  }$ and $  \mathcal{L}_{m \pm } $   gauge invariant,
 so,  $ \mathcal{L}_{z \pm}$ is also 
gauge invariant, $\delta  \mathcal{L}_{z + } =0$. Thus, if we start from open M2-branes action, we get an open $D2$-brane action with an
additional term which is gauge invariant and supersymmetric by itself.

\section{Conclusion}
In this paper we analysed two M2-branes ending on two M5-branes. 
This system was described by a BLG theory on a boundary. So, 
in this paper we analyse BLG theory in $\mathcal{N} =1$ 
superspace formalism on a manifold with two boundaries. 
 We also studied the superspace deformation 
of this theory. This deformation was expected to be caused 
by the coupling of this theory to graviphoton       background. 
It was found that the resultant theory could be made gauge and supersymmetric invariant by adding 
new boundary degrees of freedom. However \cite{faizal1}, the resultant theory only preserved half the supersymmetry 
of the original theory. We performed the Higgsing of this theory to the theory of D2-branes on a manifold with boundaries. 
The D2-brane action thus obtained was deformed by a graviphoton background. Thus, we could conclude that the theory we studied 
was dual to the $II$ string theory on a graviphoton background. Furthermore, the bulk and boundary theories where both individually 
gauge invariant after Higgsing of the theory. 

It will be interesting to 
 analyse 
the BRST and anti-BRST symmetries of this resultant theory. 
The BRST and anti-BRST symmetries of ABJM theory has already been 
studied \cite{abm}.
It will also be interesting to analyse these symmetries 
in non-linear gauges like the Curci-Ferrari gauge. 
It is expected that in this gauge these symmetries 
along with $FP$-conjugation forms the  Nakanishi-Ojima Algebra. 
This algebra is  broken due to ghost  condensation in conventional 
gauge theories. So, it will be interesting  to analyse
 if a similar thing happens for this deformed BLG theory. 
The BRST and anti-BRST symmetries of the BLG deformed by a graviphoton      
 background can also be performed.
 We can also analyse the deformation of this theory by imposing a non-anticommutative deformation between the 
fermonic coordinates. This theory will be dual to some curved  background supergravity theory.
 It will be interesting to analyse the BLG theory  dual to this curved background supergravity theory. 

In background field method all the fields of the theory are shifted. An elegant way of dealing with the BRST and the anti-BRST 
symmetries of a theory, after  shifting all the fields, is called the Batalin-Vilkovisky (BV) formalism \cite{bv}-\cite{bv1}. 
In this formalism the first the field content 
of the theory is doubled and then the Lagrangian density is chosen in such a way that along with it being
 invariant under  the original BRST and the original anti-BRST transformations, it is also invariant under these new shift transformations. 
 It is possible to express the Lagrangian density for gauge theories 
 elegantly in the 
Batalin-Vilkovisky (BV) formalism using extended superspace \cite{mf1z}. This work has also been applied to higher derivative theories
\cite{mf2z}. It will be interesting to 
perform a similar analysis for the present theory. 
Furthermore, for any gauge theory the Fock space defined in a particular gauge is  different
from those in other gauges. This is because the Fock space defined in a particular
gauge is not wide enough to realize the quantum gauge freedom. However, there is a formalism called the 
the gaugeon formalism in which it is possible to consider  quantum gauge transformation by introducing a set of
extra fields called gaugeon fields \cite{qgo}-\cite{q1go}. As the  BLG theory has a gauge symmetry associated with it, it will 
be interesting to analyse it in gaugeon formalism. It is also possible to analyse the Higgs mechanism in gaugeon formalism \cite{higgs}. Thus, we 
can also analyse the Higgs mechanism of this deformed BLG theory in gaugeon formalism.
\section{Appendix}
In this appendix we show that the vector  covariant derivative can be expressed in terms of the spinor covariant derivative. 
So, first we define  the vector covariant derivatives as 
\begin{eqnarray}
  \nabla_{ab}  X^I_A&=&  (\gamma^\mu\partial_\mu)_{ab} X_A 
-i f^{BCD}_A \Gamma_{ab BC} X^{I}_{ D},\nonumber\\
 \nabla_{ab} X^{I \dagger}_A &=&  (\gamma^\mu\partial_\mu)_{ab} X^{I \dagger}_A 
+ if^{BCD}_A X^{I \dagger}_D \Gamma^{ab BC}, \\
( \nabla_{ab} \Gamma_{de})_{AB} &=&  (\gamma^\mu\partial_\mu)_{ab} \Gamma_{de AB} 
+ C^{CD,EF}_{AB}\Gamma_{CD ab} \Gamma_{de EF}. 
\end{eqnarray}
We know that the spinor  covariant derivatives of these fields are given by 
\begin{eqnarray}
  \nabla_a  X^I_A&=& D_a X_A -i f^{BCD}_A \Gamma_{a BC} X^{I}_{ D},\nonumber\\
 \nabla_a X^{I \dagger}_A &=& D_a X^{I \dagger}_A + if^{BCD}_A X^{I \dagger}_D \Gamma^{a BC}, \\
( \nabla_a \Gamma_b)_{AB} &=& D_a \Gamma_{b AB} + C^{CD,EF}_{AB}\Gamma_{CD a} \Gamma_{b EF}. 
\end{eqnarray} 
 So, we can write 
\begin{eqnarray}
 (\{\nabla_a , \nabla_b\} X^I)_{A} &=& ( \nabla_a \nabla_b X^I)_A + ( \nabla_b \nabla_a X^I)_A \nonumber \\ 
&=& (D_a \delta^D_A -  i f^{BCD}_A \Gamma_{a CD})\nonumber \\ && \times( D_b \delta^G_D - i f^{EFG}_D \Gamma_{b EF}) X_G^I \nonumber \\ 
&& - (D_b \delta^D_A -  i f^{BCD}_A \Gamma_{b CD})\nonumber \\ && \times( D_a \delta^G_D - i f^{EFG}_D \Gamma_{a EF}) X_G^I \nonumber \\
&=&2 (\gamma^\mu\partial_\mu)_{ab} X_A 
-2i f^{BCD}_A \Gamma_{ab BC} X^{I}_{ D}
\nonumber \\ 
&=& 2 (\nabla_{ab} X^I)_A,
\nonumber \\
 (\{  \nabla_a ,   \nabla_b\} X^{I\dagger})_{A} &=& (  \nabla_a  \nabla_b X^I)_A + 
(  \nabla_b  \nabla_a X^{I\dagger})_A \nonumber \\ 
&=& (D_a \delta^D_A +  i f^{BCD}_A \Gamma_{a CD})\nonumber \\ && \times( D_b \delta^G_D + i f^{EFG}_D \Gamma_{b EF}) X_G^{I\dagger} \nonumber \\ 
&& - (D_b \delta^D_A +  i f^{BCD}_A \Gamma_{b CD})\nonumber \\ && \times( D_a \delta^G_D + i f^{EFG}_D \Gamma_{a EF}) X_G^{I \dagger} \nonumber \\ 
&=&
2 (\gamma^\mu\partial_\mu)_{ab} X^{I \dagger}_A 
+2i f^{BCD}_A X^{I \dagger}_D \Gamma_{ab BC}
\nonumber \\ 
&=& 2 (\nabla_{ab}  X^{I\dagger})_A.
\nonumber \\ 
 (\{ \nabla_a ,  \nabla_b\} \Gamma_c)_{AB} &=& ( \nabla_a  \nabla_b \Gamma_c )_{AB} + 
(  \nabla_b  \nabla_a \Gamma_c)_{AB} \nonumber \\ 
&=& (D_a \delta^A_E \delta^E_F +  C^{CD, EF}_{AB} \Gamma_{a CD})
 \nonumber \\ && \times( D_b \delta^L_E \delta^M_F +  C^{GH, LM}_{EF} \Gamma_{b GH})\Gamma_{c LM} \nonumber \\ 
&& - (D_b \delta^A_E \delta^E_F +  C^{CD, EF}_{AB} \Gamma_{b CD})\nonumber \\ && \times( D_a \delta^L_E \delta^M_F +  C^{GH, LM}_{EF} 
\Gamma_{a GH})\Gamma_{c LM} \nonumber \\ &=&
2(\gamma^\mu\partial_\mu)_{ab} \Gamma_{c AB} 
+2 C^{CD,EF}_{AB}\Gamma_{CD ab} \Gamma_{c EF}
\nonumber \\
&=& 2  (\nabla_{ab}  \Gamma_c)_{AB}.
\end{eqnarray}
Thus, we get 
\begin{equation}
  \Gamma_{ab AB} = -\frac{i}{2} \left[ D_{(a}\Gamma_{b) AB} 
- 2 i C^{CD,EF}_{AB}\{\Gamma_{a CD}, \Gamma_{b EF}\} \right].
\end{equation}

In this appendix we also show that the  covariant divergence of $\omega_{a AB}$ vanishes, 
\begin{eqnarray}
 \nabla^a \omega_{aAB} &=&[D^a \delta^E_A \delta^F_B + C_{AB}^{CD, EF}\Gamma^a_{CD}] \omega_{a EF} \nonumber \\ &=&  
-i  C^{CD,LM}_{EF}\delta^E_A \delta^F_B D^a[\Gamma^b_{CD} , D_b \Gamma_{a LM}] 
\nonumber \\ && -
 \frac{1}{3}C^{CD,LM}_{EF} C^{GH,IJ}_{LM} \delta^E_A \delta^F_B D^a[ \Gamma^b_{CD} ,
\{ \Gamma_{b GH} , \Gamma_{a IJ}\} ] 
\nonumber \\ &&
 -i  C_{AB}^{CD, EF}C^{IJ,LM}_{EF}\Gamma^a_{CD}[\Gamma^b_{IJ} , D_b \Gamma_{a LM}] 
\nonumber \\ &&  -
 \frac{1}{3}C^{CD,LM}_{EF} C^{GH,IJ}_{LM}C_{AB}^{ST, EF}
\Gamma^a_{CD}[ \Gamma^b_{ST} ,
\{ \Gamma_{b GH} , \Gamma_{a IJ}\} ] \nonumber \\ && + 
\frac{1}{2} C_{AB}^{CD, EF}\Gamma^a_{CD} D^b D_a \Gamma_{b EF} 
+   \frac{1}{2} \delta^E_A \delta^F_B D^a D^b D_a \Gamma_{b EF}
 \nonumber \\ &=& 
 0. 
\end{eqnarray}

\end{document}